\documentclass{autart}
\usepackage{natbib}
\usepackage{amsmath,amssymb,theorem}
\usepackage{graphicx}
\usepackage{algorithm,algorithmic}
\usepackage{psfrag}
\usepackage{epstopdf}
\usepackage{color,xspace}
\usepackage{subfigure}
\usepackage{caption}

\newtheorem{theorem}{Theorem}[section]

\newtheorem{remark}[theorem]{Remark}

\newtheorem{proposition}[theorem]{Proposition}

\newcommand{\real}{{\mathbb{R}}}
\newcommand{\realpositive}{\mathbb{R}_{>0}}
\newcommand{\realnonnegative}{\mathbb{R}_{\ge 0}}
\newcommand{\integernonnegative}{\mathbb{Z}_{\ge 0}}

\newcommand{\Tcap}{T_{\text{cap}}}
\newcommand{\nmax}{n_{\text{max}}}

\newcommand{\phik}{\phi_{\nu}^k}

\newcommand{\phikg}{\phi_{\gamma,\nu}^k}
\newcommand{\phikgk}{\phi_{\gamma_k,\nu}^k}
\newcommand{\phikb}{\phi_{\beta,\nu}^k}
\newcommand{\hphikg}{\hat{\phi}_{\gamma,\nu}^k}
\newcommand{\hphiklg}{\hat{\phi}_{\gamma,\nu}^{k-1}}

\newcommand{\hphikpg}{\hat{\phi}_{\gamma,\nu}^{k-l}}

\newcommand{\bphikb}{{\bar{\phi}_{\beta,\nu}}^k}

\renewcommand{\epsilon}{\varepsilon}

\newcommand{\argmax}{\operatorname{argmax}}

\renewcommand{\hat}{\widehat}

\newcommand{\oprocendsymbol}{\hbox{$\bullet$}}
\newcommand{\oprocend}{\relax\ifmmode\else\unskip\hfill\fi\oprocendsymbol}

\renewcommand{\theenumi}{(\roman{enumi})}

\begin{document}

\runauthor{Saad A. Aleem, Cameron Nowzari, and George J. Pappas}

\begin{frontmatter}

  \title{Self-triggered Pursuit of a Single Evader with Uncertain Information \thanksref{footnoteinfo}}
  \thanks[footnoteinfo]{The material in this paper was partially presented at the $54^{th}$ IEEE Conference on Decision and Control, December $15$ - $18$, $2015$, Osaka, Japan. Corresponding author: Saad A. Aleem. Tel.: +1-832-670-9958.}

  \author[address_1]{Saad A. Aleem}
  \ead{aleems@seas.upenn.edu}\qquad
  \author[address_1]{Cameron Nowzari}
  \ead{cnowzari@seas.upenn.edu}\qquad
  \author[address_1]{George J. Pappas}
  \ead{pappasg@seas.upenn.edu}

  \address[address_1]{Department of Electrical \& Systems Engineering, University of Pennsylvania, Philadelphia, PA 19104, USA}
\begin{abstract}
This paper studies a pursuit-evasion problem involving a single pursuer and a single evader, where we are interested in developing a pursuit strategy that doesn't require continuous, or even periodic, information about the position of the evader. We propose a self-triggered control strategy that allows the pursuer to sample the evader's position autonomously, while satisfying desired performance metric of evader capture. The work in this paper builds on the previously proposed self-triggered pursuit strategy which guarantees capture of the evader in finite time with a finite number of evader samples. However, this algorithm relied on the unrealistic assumption that the evader's exact position was available to the pursuer. Instead, we extend our previous framework to develop an algorithm which allows for uncertainties in sampling the information about the evader, and derive tolerable upper-bounds on the error such that the pursuer can guarantee capture of the evader. In addition, we outline the advantages of retaining the evader's history in improving the current estimate of the true location of the evader that can be used to capture the evader with even less samples. Our approach is in sharp contrast to the existing works in literature and our results ensure capture without sacrificing any performance in terms of guaranteed time-to-capture, as compared to classic algorithms that assume continuous availability of information.
\end{abstract}

\end{frontmatter}
\maketitle

\section{Introduction}\label{se:intro}
In this paper we study a continuous-time pursuit-evasion problem, involving a single pursuer and a single evader where the objective of the pursuer is to catch the evader. Traditionally, treatment of this problem assumes continuous or periodic availability of sensing/communication on the part of the agents, which entails numerous unwanted drawbacks like increased energy expenditure in terms of sensing requirement, network congestion, inefficient bandwidth utilization, increased risk of exposure to adversarial detection, etc. In contrast, we are interested in the scenario where we can relax this continuous/periodic sensing requirement for the pursuer and replace it with triggered decision making, where the pursuer autonomously
decides when it needs to sense the evader and update its trajectory to guarantee capture of the evader.
\paragraph*{Literature review:} There are two main areas related to the contents of this paper. The first is the popular problem of pursuit-evasion which has garnered a lot of interest in the past. From an engineering perspective, pursuit-evasion problems have been studied extensively in context of differential games~\citep{RI:99,TB-GJO:99}. In~\citep{JS:01}, sufficient conditions are derived for a pursuer to capture an evader where the agents have equal maximum speeds and are constrained to move within the nonnegative quadrant of $\mathbb{R}^2$. In~\citep{LA-ASG-EMR:92}, upper and lower bounds on the time-to-capture have been discussed where the agents are constrained in a circular environment. These pursuit strategies have been generalized and extended by~\citep{SK-CVR:05} to guarantee capture using multiple pursuers in an unbounded environment $\mathbb{R}^n$, as long as the evader is initially located inside the convex hull of the pursuers. In context of multi-agent robotic systems, the visibility-based pursuit-evasion has received a lot of interest in the past~\citep{SML-JEH:01,SS-SML-SR:04,VI-SK-SK:05}. In these problems, the pursuer is visually searching for an unpredictable evader that can move arbitrarily fast in a simply connected polygonal environment. Similar problems have been studied in~\citep{IS-MY:92,BPG-ST-GG:06,VI-SK-SK:06}, where visibility limitations are introduced for the pursuers but the agents actively sense and communicate at all times. A related problem has been discussed in~\citep{SDB-FB-JH:07a}, where the agents can move in $\mathbb{R}^2$ but each agent has limited range of spatial sensing. A detailed review of recent applications in context of search and rescue missions and motion planning involving adversarial elements can be found in~\citep{THC-GAH-VI:11}.

The vast literature on pursuit-evasion problems highlights their multifaceted applications in a variety of contexts. However, the previous works usually assume continuous or periodic availability of sensing information, especially on the part of the pursuer. Towards this end, we want to apply the new ideas of triggered control to the problem of pursuit-evasion, which has not been studied so far. In contrast to conventional time-driven approaches, strategies based on triggered control schemes study how information could be sampled for control purposes where the agents act in an opportunistic fashion to meet their desired objective~\citep{KJA-BB:99,MV-JF-PM:03}. Triggered control allows us to analyze the cost to make up for less communication effort on the part of the agents, while achieving a desired task with a guaranteed level of performance of the system (see~\citep{WMPHH-KHJ-PT:12} for an overview of more recent studies). Of particular relevance to this paper are works that study self-triggered~\citep{RS-FF:06,CN-JC:12} or event-triggered~\citep{AE-DVD-KJK:10,MM-PT:11} implementations of local agent strategies. In event-triggered control, the focus is on detecting events (both intrinsic and exogenous) during the execution that trigger pre-defined agent actions. In self-triggered control, the emphasis is instead on developing autonomous tests that rely only on current information available to individual agents to schedule or pre-compute future actions. In the context of pursuit-evasion problems, we will make use of the self-triggered approach which equips the pursuer with autonomous decision making in order to decrease its required sensing effort in tracking the evader. In principle, our paper shares with the above works the aim of trading increased decision making at the agent (pursuer) level for less sensing effort while still guaranteeing capture of the evader.
\paragraph*{Contribution:} This paper builds on our earlier work in~\citep{SAA-CN-GJP:15}, where we applied the framework of triggered control to design a self-triggered pursuit policy for the pursuer which guarantees capture of the evader with a finite number of observations. Our work was different from the existing methods in the literature as our analysis did not assume the availability of continuous or periodic information about the evader. Instead, the self-triggered framework guaranteed capture of the evader without sacrificing any performance, in terms of guaranteed time-to-capture, as compared to classic algorithms that assume continuous information is available at all times.

The key result in~\citep{SAA-CN-GJP:15} relied on receiving perfect information about the evader, whenever the pursuer decided to sample. In reality, exact position information about the evader may never be available to the pursuer. The main contribution in this paper is to design a robust self-triggered policy, where we allow for noisy sensor measurements on the part of the pursuer, while still guaranteeing capture with only sporadic evader observations. Our triggered control framework provides fresh insights into dealing with information uncertainties and worst-case scenarios in the pursuit-evasion problem. Our framework readily incorporates the uncertainty in sensing the evader and allows us to derive tolerable error bounds in estimating the evader's position that preserve our capture guarantees. The theme of our pursuit policy is quite similar to existing works on triggered control; based on the latest current estimate of the evader, the pursuer computes a certificate of sleep duration for which it can follow its current trajectory without having to sense the evader. In addition, we discuss the relative advantages of retaining previous estimates, where we leverage the past information about the evader to arrive at a better estimate of the evader's true location. We show that incorporating additional knowledge about evader's past improves the self-triggered update duration for the pursuer and mitigates uncertainty in detecting the evader.
\paragraph*{Organization:} The problem formulation and its mathematical model are presented in Section~\ref{se:statement}. In Section~\ref{se:self}, we present the design of self-triggered update duration for the pursuer as derived in~\citep{SAA-CN-GJP:15}. It is followed by Section~\ref{se:error}, where we allow for uncertainty in sampling the evader's position and outline the maximum tolerable error that can be accommodated on the part of the pursuer, without compromising its self-triggered strategy. Section~\ref{se:prev_estimates} discusses the relative merits of retaining previous estimates of the evader's position in the hope of increasing the sleep durations for the pursuer. The readers are encouraged to go over the detailed analysis of our problem in the Appendix.
\paragraph*{Notation:} We let $\realpositive$, $\realnonnegative$  and $\integernonnegative$ to be the sets of positive real, nonnegative real and nonnegative integer numbers, respectively. $\real^n$ denotes the $n-$dimensional Euclidean space and $\| \cdot \|$ is the Euclidean distance.

\section{Problem statement}\label{se:statement}
We consider a system with a single pursuer $P$ and a single evader $E$. At any given time $t$, the position of the evader is given by $r_e(t) \in \real^2$ and its velocity is given by $u_e(t) \in \real^2$ with $\| u_e(t) \| \leq v_e$, where $v_e$ is the maximum speed of the evader. Similarly, the position and velocity of the pursuer are given by $r_p(t)$ and $u_p(t)$ with $\| u_p(t) \| \leq v_p$, where $v_p > v_e$ is the maximum speed of the pursuer. The system evolves as
\begin{align*}
\dot{r}_p &= u_p,\\
\dot{r}_e &= u_e.
\end{align*}

In our problem, the goal of the pursuer is to capture the evader. We define \textit{capture} of the evader as the instance when the pursuer is within some pre-defined capture radius $\epsilon > 0$ of the evader. Assuming that the pursuer has exact information about the evader's state at all times, it is well known that the time-optimal strategy for the pursuer is to move with maximum speed in the direction of the evader~\citep{RI:99}. Such a strategy, known as classical pursuit, is given by the control law
\begin{align}\label{eq:p_cl_classic}
u_p(t) &=  v_p\frac{r_e(t) - r_p(t)}{\|r_e(t) - r_p(t)\|}.
\end{align}

The issue with the control law~\eqref{eq:p_cl_classic} is that it requires continuous access to the evader's
state at all times and instantaneous updates of the control input. Instead, we want to guarantee capture of the evader without tracking it at all times and only updating the controller sporadically. We do this by having the pursuer decide in an opportunistic fashion when to sample the evader's position, and update its control input. Under this framework, the pursuer only knows the position of the evader at the time of its last observation.
Let $\{ t_k \}_{k \in \integernonnegative}$ be a sequence of times at which
the pursuer receives information about the evader's position. In between updates, the pursuer implements a zero-order hold of the control signal computed at the last time of observation using~\eqref{eq:p_cl_classic} which is given by
\begin{align} \label{eq:p_cl_self}
u_p(t) &= v_p \frac{r_e(t_k) - r_p(t_k)}{\|r_e(t_k) - r_p(t_k)\|},
\end{align}
for $t \in [t_k, t_{k+1})$.

In this paper, our purpose is to identify a function for the self-triggered update duration $\phi$
for the pursuer that determines the next time at which the updated information is required. In other words,
each time the pursuer receives updated information about the evader at some time $t_k$, we want
to find the duration, $\phi\left(D_k, v_e, v_p\right)$, until the next update such that
\begin{align}\label{eq:p_self_update}
   t_{k+1}  = t_k + \phi\left(D_k, v_e, v_p\right) ,
\end{align}
where $D_k \triangleq \|r_e(t_k) - r_p(t_k)\|$ is the separation between the agents at time $t_k$.
Our goal is to design the triggering function~$\phi$ such that the pursuer is guaranteed to capture the
evader while also being aware of the number of samples of the evader required.

\section{Design of Self-triggered Update Law}\label{se:self}
We study the continuous-time pursuit and evasion problem consisting of a single pursuer and a single evader on a plane ($\real^2$), where both agents are modelled as single integrators. Note that it is not necessary to assume that the agents, particularly the evader, are moving with constant speeds at all times. For all practical purposes, we can upper bound the evader speed by $v_e^{\max}$ such that $v_e^{\max} < v_p$ and the analysis will remain unchanged.

\begin{figure}[!h]
\centering
\includegraphics[scale =.6]{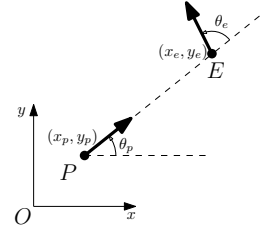}
\caption{Figure shows the pursuer $P$ at $r_p = (x_p,y_p)$ and the evader $E$ at $r_e = (x_e,y_e)$ in $\real^2$. The pursuer is moving along $\theta_p$ and the relative angle between agents' headings is denoted by $\theta_e$. The arrows indicate the velocity vectors of the agents.}
\label{fig:pe_geometry}
\end{figure}

We denote the positions of the agents by $r_p = \left(x_p,y_p\right)$ and $r_e = (x_e,y_e)$. Additionally, the pursuer is moving along $\theta_p$ and the relative angle between the agents' headings is denoted by $\theta_e$ (see Fig.~\ref{fig:pe_geometry}). Without loss of generality, we normalize the speed of the pursuer to $v_p = 1$ and the evader moves with a speed $\nu$, where $\nu < 1$ at all times. The dynamics of the pursuer and the evader are given by
\begin{equation}\label{eq:pe_dyn}
\begin{aligned}
\dot{x}_p &= \cos \theta_p, &\dot{x_e} &= \nu \cos (\theta_e + \theta_p),\\
\dot{y}_p &= \sin \theta_p, &\dot{y_e} &= \nu \sin (\theta_e + \theta_p).
\end{aligned}
\end{equation}

\subsection{Self-triggered Update Policy for Pursuer}
Suppose at time $t_k$ the pursuer, at $r_p(t_k)$, observes the evader at $r_e (t_k)$ such that the distance between the agents is $ D_k \triangleq\|r_p(t_k) - r_e(t_k)\|$. For notational brevity, we will denote the position of the agents at the instance of observation by $r_e^k:=r_e(t_k)$ and $r_p^k:=r_p(t_k)$. We are interested in the duration for which the pursuer can maintain its course of trajectory, without observing the evader. More specifically, we are interested in the first instance at which the separation between the agents can possibly increase, thus prompting the pursuer to sample the evader's state and update its trajectory. Let $r(t)$ denote the separation between the pursuer and the evader at time $t$. Then, we consider the objective function
\[ R = \frac{r^2}{2}  = \frac{(x_e-x_p)^2 + (y_e-y_p)^2}{2} .\]
Note that the time at which $\dot{R}$ becomes nonnegative is same as the time at which $\dot{r}$ becomes nonnegative. Using~\eqref{eq:pe_dyn}, the derivative of $R$ (see Appendix for details) is given by
\begin{align}\label{eq:dot_R_all}
\dot{R}&= \nu(x_e - \tau) \cos \theta_e + \nu y_e \sin \theta_e + \tau - x_e,
\end{align}
for $\tau \ge 0$, and is a function of evader parameters $(x_e, y_e, \theta_e)$. For $\tau \ge 0$, the reachable set of the evader is given by the ball $B_e(r_e^k,\nu\tau)$. Additionally, for fixed $\tau$, we write $\dot{R}$ in~\eqref{eq:dot_R_all} explicitly as a function of evader parameters and denote it by $\dot{R}_{\tau}(x_e,y_e,\theta_e)$. Let $g(\tau) = \underset{x_e, y_e, \theta_e}{\sup} \dot{R}_{\tau}(x_e,y_e,\theta_e)$, subject to the reachable set of the evader, i.e $r_e \in B_e(r_e^k,\nu\tau)$. For the dynamics in~\eqref{eq:pe_dyn}, where $v_p = 1$ and $v_e = \nu$, we can denote update duration in~\eqref{eq:p_self_update} by $\phik \triangleq \phi(D_k,v_p,v_e)$ and it is defined as
\[\phik = \inf \left\{\tau \in \realpositive \,|\, g(\tau) = 0 \right\}.\]
$\phik$ is the smallest duration after which there exists an evader state that may increase the separation between the agents.
For the agents modelled by~\eqref{eq:pe_dyn}, our self-triggered update duration is obtained by solving $\inf \left\{\tau \in \realpositive \,|\, g(\tau) = 0 \right\}$ (see Appendix for derivation) and is given by
\begin{align}\label{eq:t_min_v}
\phik&= \frac{D_k\nu \sqrt{1 - \nu^2} - D_k(1 - \nu^2)}{2\nu^2 - 1}.
\end{align}
{\psfrag{fourfivesix1}[cc][cc]{\small $\nu$}
\psfrag{fourfivesix2}[cc][cc]{\tiny $\frac{\phik}{D_k}$}
\begin{figure}
\centering
\includegraphics[scale = 0.4]{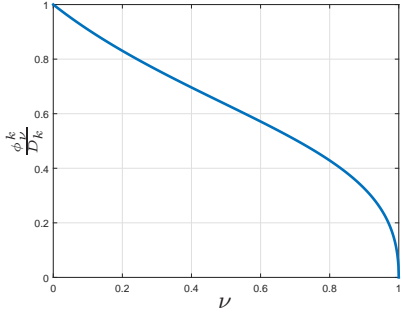}
\caption{Figure shows a plot of normalized update time $\frac{\phik}{D_k}$ against evader speed $\nu \in [0, 1)$. $\phik(D_k)$ is given by~\eqref{eq:t_min_v}.}
\label{fig:t_min_v}
\end{figure}}
The graph of $\phik$ against evader speed $\nu$ is shown in Fig.~\ref{fig:t_min_v}. From the plot, we observe that increasing the evader speed decreases the self-triggered update duration for the pursuer. So, if the evader moves faster our law prescribes more frequent updates of it to guarantee capture.

The underlying objective in the design of the self-triggered policy is that at each instance of fresh observation, the separation between the agents must have decreased. The following proposition characterizes this result.
\begin{proposition}[Decreasing Separation]\label{prop:1}
Let the pursuer and evader dynamics be given by~\eqref{eq:pe_dyn} where the agents are separated by $D_k$ at time $t_k$. If the pursuer updates its trajectory using the self-triggered update policy $\phik$ in~\eqref{eq:t_min_v}, then the distance between the agents at time $t_{k+1}$ has strictly decreased, i.e., $D_{k+1} < D_k$ for $t_{k+1} = t_k + \phik$.
\end{proposition}
\begin{pf}
Given the separation $D_k$ at time $t_k$, the new separation between the agents, after a duration of $\phi_{\nu}^k$, is $D_{k+1}$. To see that the separation is strictly decreasing, note that
\begin{align*}
D_{k+1} \le D_{\max}^k = D_k - (1 - \nu)\phik (D_k) \triangleq D_k h(\nu),
\end{align*}
where $D_{\max}^k$ is the maximum possible separation between the agents after the duration $\phik$ (see Appendix for details) and $h(\nu)$ is given by
\begin{align}\label{eq:h_nu}
h(\nu) &= 1 - \frac{\nu(1 - \nu) \sqrt{1 - \nu^2} - (1 - \nu)(1 - \nu^2)}{2\nu^2 - 1},
\end{align}
where $h(\nu) \in [0,1)$ for $\nu \in [0,1)$ . Thus, $D_{k+1} < D_k$. \qed 
\end{pf}
\subsection{Capture Time \& Number of Samples}
Using the self-triggered update policy~\eqref{eq:t_min_v}, the pursuer is guaranteed to capture the evader in finite time with finite number updates. More specifically, we can find the maximum number of samples in terms of the capture radius $\epsilon$ and evader speed $\nu$ and use it to guarantee finite time-to-capture. This is summarized in the following theorem.
\begin{theorem}[Capture with Finite Samples]\label{thm:1}
Let the pursuer and evader dynamics be given by~\eqref{eq:pe_dyn} where the agents are initially separated by $D_0$. Given some pre-defined positive capture radius $\epsilon < D_0$, the self-triggered update policy $\phik$ in~\eqref{eq:t_min_v} ensures capture with finite observations in finite time.
\end{theorem}
\begin{pf} 
According to Proposition~\ref{prop:1}, the separation between the agents is strictly decreasing between successive updates. In fact, the new separation between the agents satisfies the inequality $D_{k+1} \le D_{k} h(\nu) $, where $h(\nu)$ is given by~\eqref{eq:h_nu}. This implies that after $n$ observations of the evader, the separation between the agents satisfies the inequality $D_n \le D_0 h^n(\nu)$, where $D_0$ is the initial separation between the agents. Using this result, the maximum number of samples can be calculated by setting $D_0h^n(\nu) \le \epsilon$.
\begin{align}\label{eq:n_samples}
\nmax &= \left\lceil\frac{\log\left(\frac{\epsilon}{D_0}\right)}{\log\left(h(\nu)\right)} \right\rceil.
\end{align}
The expression in~\eqref{eq:n_samples} shows that for any pre-defined positive capture radius $\epsilon < D_0$, the pursuer is guaranteed to capture the evader with \emph{finite} number of samples. This completes the first part of the proof. For self-triggered pursuit policy $\phik$ in~\eqref{eq:t_min_v}, the sequence of times at which the pursuer samples the evader position, denoted by $\{t_k\}_{k \in \integernonnegative}$, follows the criteria $t_{k+1} = t_k + \phik$. This means that after $N$ updates, the total duration of pursuit (denoted  by $T_N$) is given by $ T_N = t_0 + \sum_{k = 0}^N \phik$. Without loss of generality, we can assume $t_0 = 0$. Since the pursuer is guaranteed to capture the evader with finite number of maximum samples $\nmax$, the time-to-capture (denoted by $\Tcap$) is bounded by \[\Tcap \le \sum_{k = 0}^{\nmax} \phik = f(\nu)\sum_{k = 0}^{\nmax}D_k,\]
where $f(\nu)$ denotes $\frac{\phik}{D_k}$ 
and satisfies the relationship $f(\nu) = \frac{1 - h(\nu)}{1 - \nu}$. Using the inequality $D_k \le D_0 h^k(\nu)$, we get
\[\Tcap \le D_0 f(\nu) \sum_{k = 0}^{\nmax} h^k(\nu).\] As $h(\nu) \in [0,1)$ for $\nu \in [0,1)$, we have
  \begin{align*}
  \sum_{k = 0}^{\nmax} h^k(\nu) <\sum_{k = 0}^{\infty} h^k(\nu)&=  \frac{1}{1 - h(\nu)}.
 \end{align*}
Thus, $\Tcap < \frac{D_0}{1 - \nu}$, where in the last step we have made use of the relationship $f(\nu) = \frac{1 - h(\nu)}{1 - \nu}$. This shows that, for any evader speed $\nu \in [0,1)$, $\Tcap$ is finite. \qed
\end{pf}

\begin{remark}[Self-triggered Performance]
{\rm Note that in proving finite time-to-capture in Theorem~\ref{thm:1}, we showed that time-to-capture is strictly less than $\frac{D_0}{1 - \nu}$. However, given a pre-defined capture radius $\epsilon < D_0$, it can be shown that the time-to-capture satisfies the inequality
\begin{align}\label{eq:time_capture}
\Tcap &\le \frac{D_0 - \epsilon}{1 - \nu}.
\end{align}
This is because the evader is captured as soon as the actual separation is within the capture radius at any time, not just at the instance of updates. The relationship in~\eqref{eq:time_capture} is the same upper bound for time-to-capture in classical pursuit strategy. In classical pursuit, 
$\Tcap$ is bounded by
\begin{align} \label{eq:time_capture_classic}
\Tcap & \le \frac{D_0 - \epsilon}{v_p - v_e},
\end{align}
where $D_0$ is the initial separation,  $\epsilon$ is the pre-defined capture radius and $v_p>v_e$~\citep{RI:99}. The worst-case time-to-capture occurs in the scenario where evader is actively moving away at all times. This shows that our self-triggered pursuit policy guarantees capture with the \emph{same} performance as the classical case but with only finite number of evader samples.}\oprocend
\end{remark}

The expression in~\eqref{eq:n_samples} guarantees capture with finite samples of evader's state. Fig.~\ref{fig:n_samples} shows a graph of the maximum number of samples required to guarantee capture against the evader speed, for pre-defined capture radius $\epsilon = \frac{D_0}{10^3}$. The number of samples increase quite sharply as $\nu$ approaches $1$. This makes intuitive sense as the maximum number of evader observations should increase as evader approaches the maximum speed of the pursuer.
{
\psfrag{fourfivesix1}[cc][cc]{\small $\nu$}
\psfrag{fourfivesix2}[cc][cc]{\tiny Maximum number of samples}
\begin{figure}
\centering
\includegraphics[scale = 0.4]{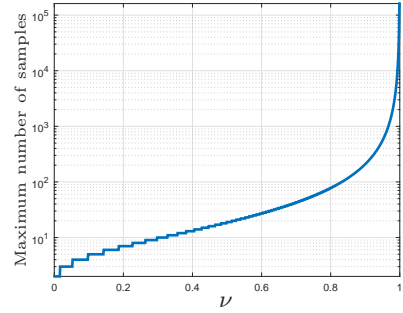}
\caption{Figure shows the variation of the maximum number of samples against evader speed as given by the expression in~\eqref{eq:n_samples} where $\epsilon$ is chosen as $\frac{D_0}{10^3}$.}
\label{fig:n_samples}
\end{figure}
}

\section{ Allowable Error in Evader's Estimate}\label{se:error}
In this section, we study the scenario in which the pursuer acquires the information about the position of the evader with some uncertainty. We are interested in analyzing the effect of imperfect observations on our self-triggered framework. Our objective is to investigate the maximum allowable error in estimating the evader's position which still allows us to catch the evader using a self-triggered pursuit policy. More specifically, we want to find the maximum tolerable uncertainty in estimating evader's position at each instance of observation, as a function of the evader's speed.

\begin{figure}[!h]
\centering
\includegraphics[scale = 0.5]{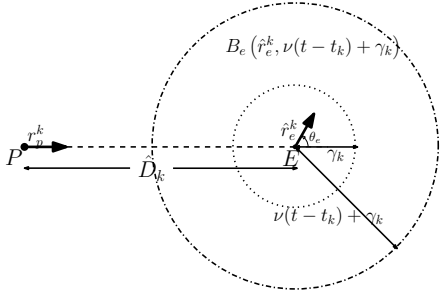}
\caption{Figure shows the pursuer at $r_p^k$ estimating the observer at $\hat{r}_e^k$, separated by $\hat{D}_k$, at time $t_k$. The pursuer detects the evader with an uncertainty $\gamma_k$. $B_e\left(\hat{r}_e^k, \nu (t-t_k) + \gamma_k\right)$ indicates the reachable set of the evader for $t \in [t_k, t_{k+1})$.}
\label{fig:pe_analysis_error}
\end{figure}

\begin{figure}[!h]
\centering
\includegraphics[scale = 0.65]{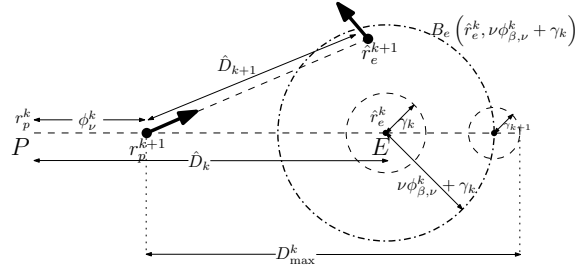}
\caption{At time $t_k$, the pursuer $P$, measures the evader $E$ at $\hat{r}_e^k$. After the duration $\phikb$, $\hat{D}_{k+1}$ indicates $\|r_p^{k+1} - \hat{r}_e^{k+1}\|$. $B_e(\hat{r}_e^k, \nu \phikb + \gamma_k)$ outlines the boundary of the reachable set of the evader after $\phikb$. $D_{\max}^k$ denotes the maximum possible separation between the agents at $t_{k+1} = t_k + \phikb$.}
\label{fig:pe_analysis_cont_error}
\end{figure}

Suppose at $t_k$ the pursuer estimates the evader at $\hat{r}_e(t_k)$. This observation is imperfect and is corrupted by an associated noise $\gamma_k$, where $\gamma_k \in \realnonnegative$. So, at the instance of observation, the true position of the evader $r_e(t_k) \in B_e\left(\hat{r}_e(t_k), \gamma_k\right)$. For notational brevity we will denote $\hat{r}_e(t_k)$ by $\hat{r}_e^k$. Our objective is to find the allowable range for the error $\gamma_k$, as a function of the evader speed $\nu$, such that the self-triggered pursuit policy can still guarantee capture. The reachable set of the evader is given by $B_e\left(\hat{r}_e^k, \nu(t-t_k) + \gamma_k\right)$, for $t \in [t_k, t_{k+1})$. This is illustrated in Fig.~\ref{fig:pe_analysis_error}. Let $\tau = t - t_k$. Applying the previous framework of analysis, we want to maximize $\dot{R}$ over the evader parameters, subject to the constraint $r_e \in B_e(\hat{r}_e^k, \nu\tau + \gamma_k)$.
Let $g_{\gamma_k}(\tau) = \underset{x_e, y_e, \theta_e}{\sup} \dot{R}_{\tau}(x_e,y_e,\theta_e)$, subject to the reachable set of the evader, i.e. $r_e \in B_e(\hat{r}_e^k, \nu\tau + \gamma_k)$. Finding $g_{\gamma_k}(\tau)$ is very similar to the procedure of finding $g(\tau)$ as outlined in the Appendix. The only difference is in the reachable set of the evader, which is now increased by $\gamma_k$, whereas $\dot{R}$ remains the same. For the error $\gamma_k$ and estimated separation $\hat{D}_k$, the self-triggered update duration is defined as
 \[ \phikgk =\phi(\hat{D}_k, \gamma_k,\nu) \triangleq \inf \left\{\tau \in \realpositive\,|\, g_{\gamma_k}(\tau) = 0 \right\}.\]
Solving $\inf \left\{\tau \in \realpositive\,|\, g_{\gamma_k}(\tau) = 0 \right\}$ yields
\begin{align}\label{eq:self_trig_gamma}
\phikgk &= \frac{\hat{D}_k\nu \sqrt{1 - \nu^2} - \hat{D}_k(1 - \nu^2) + \gamma_k(\sqrt{1 - \nu^2} - \nu)}{2\nu^2 - 1},
\end{align}
for $\nu \in [0,1)$. To simplify the analysis, we can select the error as a scaled version of the current estimate of the separation, i.e.  $\gamma_k \triangleq \beta \hat{D}_k$, where $\beta \in \realnonnegative$. Such a parametrization will allow us to study the effect of changing $\beta$ on the self-triggered update duration and will also tell us the maximum tolerable error relative to the current estimate of the separation $\hat{D}_k$. Setting $\gamma_k := \beta \hat{D}_k$, we get
\begin{align}\label{eq:self_trig_beta}
\phikb &= \frac{\hat{D}_k\left(\nu \sqrt{1 - \nu^2} - (1 - \nu^2) + \beta(\sqrt{1 - \nu^2} - \nu)\right)}{2\nu^2 - 1},
\end{align}
where, with a slight abuse of notation, we have used $\phikb$ instead of $\phi(\hat{D}_k, \beta,\nu) = \inf \{\tau \in \realpositive\,|\, g_\beta (\tau) = 0\}$\footnote{$g_{\beta}(\tau) = \underset{x_e, y_e, \theta_e}{\sup} \dot{R}_{\tau}(x_e,y_e,\theta_e)$, s.t. $r_e \in B_e(\hat{r}_e^k, \nu\tau + \beta \hat{D}_k)$}. Note that $\beta$ can not be chosen arbitrarily. To find the feasible domain of the of $\beta$, we invoke the criteria of $\phikb \in \realpositive$ for $\nu \in [0,1)$. This yields $\beta \in \left[0, \sqrt{1 - \nu^2}\right)$. However, imposing the positivity of self-triggered duration is not sufficient to come up with the desired domain. The design of the self-triggered update duration rests on the underlying performance objective of evader capture. This requires strict decrease in true separation in between updates\footnote{In the case of uncertainty, we have access to the estimates of current separation $\hat{D}_k$, instead of true separation $D_k$}.  At the instance of update, the new estimate of the separation satisfies $\hat{D}_{k+1} \le D_{\max}^k$ where $D_{\max}^k$ is given by
\[D_{\max}^k = \hat{D}_k + \nu \phikb + \gamma_k + \gamma_{k+1} -  \phikb.\]
This is illustrated in Fig.~\ref{fig:pe_analysis_cont_error}.
Thus,
\begin{align*}
\hat{D}_{k+1} &\le D_{\max}^k= \hat{D}_k + \nu \phikb + \gamma_k + \gamma_{k+1} -  \phikb.
\end{align*}
Recall that $\phikb$ is a function of $\hat{D}_k$. Let $\bphikb \triangleq \frac{\phikb}{\hat{D}_k}$. This means that $\hat{D}_{k+1} \le \hat{D}_k \left(  \frac{1 - (1 - \nu)\bphikb + \beta}{1 - \beta} \right)$.  In deriving the previous inequality, we have allowed for the worst-case scenarios in estimating the evader's position. Therefore, setting $\frac{1 - (1 - \nu)\bphikb + \beta}{1 - \beta} < 1$ is sufficient to guarantee strict decrease in actual separation in between updates. This results in a more conservative set of allowable values for $\beta$ as shown in~\eqref{eq:beta_domain}.
\begin{equation}\label{eq:beta_domain}
\begin{aligned}
\beta \in \left[0, \frac{(1 - \nu)\sqrt{1 - \nu^2} - 2(1 - \nu)^2}{5\nu - 3}\right).
\end{aligned}
\end{equation}
{
\psfrag{fourfivesix1}[cc][cc]{\small$\beta$}
\psfrag{fourfivesix2}[cc][cc]{\small $\frac{\phikb}{D_k}$}
\psfrag{onetwothree1}[cc][cc]{\tiny $\nu = 0.2$}
\psfrag{onetwothree2}[cc][cc]{\tiny $\nu = 0.4$}
\psfrag{onetwothree3}[cc][cc]{\tiny $\nu = 0.6$}
\psfrag{onetwothree4}[cc][cc]{\tiny $\nu = 0.8$}
\begin{figure}
\centering
\includegraphics[scale = 0.4]{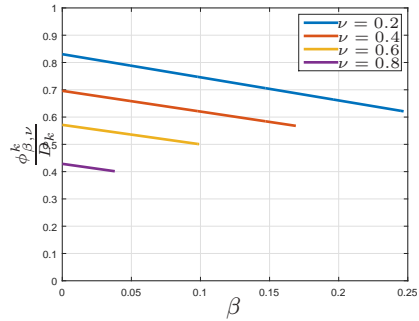}
\caption{Figure shows the variation of $\phikb$ in~\eqref{eq:self_trig_beta} against $\beta$, for different values of evader speeds, where $\beta$ satisfies the condition in~\eqref{eq:beta_domain}. For any evader speed $\nu \in[0,1)$, increasing the value of $\beta$ decreases $\phikb$.}
\label{fig:t_min_b}
\end{figure}
}
Thus, for evader speed $\nu$, the maximum allowable error for $k$-th observation (denoted by $\gamma_k^*$) satisfies the inequality
\[ \gamma^*_k < \frac{\hat{D}_k\left((1 - \nu)\sqrt{1 - \nu^2} - 2(1 - \nu)^2\right)}{5\nu - 3}.\] Note that the maximum allowable error dynamically changes (decreases) as the pursuer closes in on the evader. This also means that, for the duration of \emph{entire} pursuit, the maximum allowable error for \emph{all} observations (denoted by $\gamma^*$) satisfies the relationship
 \[ \gamma^* < \frac{\epsilon\left((1 - \nu)\sqrt{1 - \nu^2} - 2(1 - \nu)^2\right)}{5\nu - 3},\]
where $\epsilon$ is the pre-defined capture radius.

We incur no Zeno behavior using the self-triggered policy in~\eqref{eq:self_trig_beta}, i.e. the pursuer does not require infinitely many samples to capture the evader. The following theorem characterizes this important result.
\begin{theorem}[No Zeno Behavior]\label{thrm:2}
If the pursuer updates its trajectory using the self-triggered update policy $\phikb$ in~\eqref{eq:self_trig_beta}, where $\beta$ satisfies the condition in~\eqref{eq:beta_domain}, then pursuer is guaranteed to incur no Zeno behavior for the duration of the pursuit.
\end{theorem}
\begin{pf}
Suppose that pursuer observes the evader at the instance $t_k$. In order to show no Zeno behavior, it suffices to prove that the inter-event duration is lower-bounded by a positive constant for all observations, i.e. $t_{k+1} - t_k:=\phikb \ge c > 0$, $\forall k$. According the condition~\eqref{eq:beta_domain}, $\beta \in \left[0, \frac{(1 - \nu)\sqrt{1 - \nu^2} - 2(1 - \nu)^2}{5\nu - 3}\right)$ for the self-triggered update policy $\phikb$ in~\eqref{eq:self_trig_beta}. Note that the higher values of $\beta$ result in a smaller inter-event duration (see Fig.~\ref{fig:t_min_b}). Then, given a capture radius $\epsilon$, the following relationship is satisfied for all observations.
\begin{align*}
\phikb \ge \phi(\epsilon,\nu) &= \epsilon q(\nu),
\end{align*}
where $q(\nu) = \frac{\nu \sqrt{1 - \nu^2} - (1 - \nu^2) + p(\nu)(\sqrt{1 - \nu^2} - \nu)}{2\nu^2 - 1}$ and $p(\nu) = \frac{(1 - \nu)\sqrt{1 - \nu^2} - 2(1 - \nu)^2}{5\nu - 3}$. It is easy to verify that $q(\nu)$ is a positive and monotonically decreasing function for $\nu \in [0,1)$. Thus, for any evader speed $\nu \in [0,1)$, we have $t_{k+1} - t_k:=\phikb \ge \epsilon q(\nu) > 0$, $\forall k$. So, inter-event duration is lower-bounded by a positive constant for all observations, which suffices to show that our self-triggered duration does not incur any Zeno behavior.\qed
\end{pf}

An important consequence of Theorem~\ref{thrm:2} is that, in the presence of uncertainty, our self-triggered framework guarantees capture with only finite estimates of the evader. This means that we can find the maximum number of samples that will guarantee capture. By design, $\beta$ satisfies the condition in~\eqref{eq:beta_domain} and guarantees strict decrease in measured separation between successive updates. In general, we can write this as
\begin{align*}
\hat{D}_{k+1} &\le \hat{D}_k h_\beta(\nu),
\end{align*}
\addtocounter{footnote}{-2}
where $h_\beta(\nu)$\footnote{$h_{\beta}(\nu) = \frac{1 - \frac{(1 - \nu)\left(\nu \sqrt{1 - \nu^2} - (1 - \nu^2) + \beta(\sqrt{1 - \nu^2} - \nu)\right)}{2\nu^2 - 1} + \beta}{1 - \beta}.$} $\in [0,1)$ for $\nu \in [0,1)$ and $\beta$ satisfies the condition in~\eqref{eq:beta_domain}. Using similar analysis from previous section, we can estimate the maximum number of samples in the presence of uncertainty. Given an initial estimated separation of $\hat{D}_0$ between agents and a pre-defined positive capture radius $\epsilon < \hat{D}_0$, the maximum number of evader updates can be calculated by setting $\hat{D}_0 \left(h_{\beta}(\nu)\right)^n \le \epsilon$, which results in
\begin{align}\label{eq:n_samples_beta}
\nmax^{\beta} &= \left\lceil\frac{\log\left(\frac{\epsilon}{\hat{D}_0}\right)}{\log\left(h_{\beta}(\nu)\right)} \right\rceil.
\end{align}
For evader speed $\nu$, the maximum number of samples for guaranteed capture \emph{increases} as the parameter $\beta$ is increased (see Fig.~\ref{fig:n_samples_beta}). This shows that, for any $\nu \in [0,1)$, we need to sample the evader's state more frequently in order to allow for more uncertainty in estimating evader's position at each instance of update.
{
\psfrag{fourfivesix1}[cc][cc]{\small$\beta$}
\psfrag{fourfivesix2}[cc][cc]{\tiny $\text{Maximum number of samples}$}
\psfrag{onetwothree1}[cc][cc]{\tiny $\nu = 0.20$}
\psfrag{onetwothree2}[cc][cc]{\tiny $\nu = 0.40$}
\psfrag{onetwothree3}[cc][cc]{\tiny $\nu = 0.60$}
\psfrag{onetwothree4}[cc][cc]{\tiny $\nu = 0.80$}
\begin{figure}[htb]
\centering
\includegraphics[scale = 0.45]{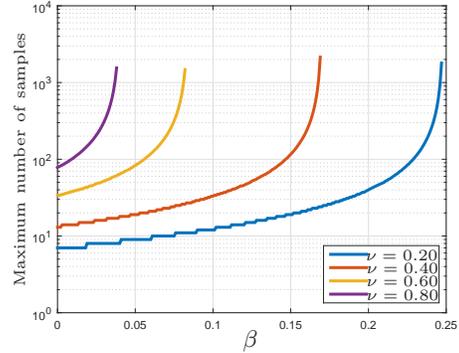}
\caption{Figure shows the plot of maximum number of samples in~\eqref{eq:n_samples_beta} against $\beta$, for different values of evader speeds, where $\beta$ satisfies the condition in~\eqref{eq:beta_domain}. The capture radius is taken as $\epsilon = \frac{\hat{D}_0}{10^3}$.}
\label{fig:n_samples_beta}
\end{figure}}

\section{Self-triggered Analysis with Memory}\label{se:prev_estimates}
In the previous section, we studied the effect of uncertainty in sensing the evader's position and provided bounds on the maximum tolerable error (as a function of evader's speed) which allowed us to capture the evader with sporadic updates. In the absence of uncertainty, as outlined in Section~\ref{se:self}, the pursuer employs a \emph{memoryless} pursuit policy to catch the evader, i.e. the pursuer only needs the \emph{current} sample of the evader's (true) position $r_e^k$ in order to calculate the self-triggered update duration $\phik$ in~\eqref{eq:t_min_v}. This is because, for $\tau \in [0,\phik)$, the \emph{current} reachable set of the evader, $B_e(r_e^k,\nu \tau)$, is always a subset of the \emph{previous} reachable set of the evader, $B_e\left(r_e^{k-1},\nu(\phik + \tau)\right)$. When we introduce uncertainty in estimating evader's position, the above statement is no longer true. The idea behind retaining evader's estimated history is that we can potentially reduce the \emph{actual} reachable set of the evader when we combine the current reachable set of the evader with the previous reachable sets, thus improving our current estimate of the true position of the evader. This allows us to improve (increase) our update duration, as compared to the memoryless case \emph{with} uncertainty (derived in Section~\ref{se:error}). More specifically, we can leverage our knowledge about the previous estimates of the evader in improving the current update duration for the pursuer, while mitigating the effect of uncertainty in estimating evader's position.

\begin{figure}[!h]
\centering
\includegraphics[scale =.5]{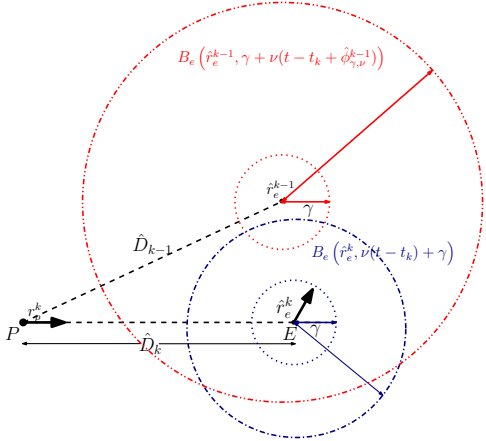}
\caption{The figure shows the reachable sets of the evader, based on its current ($\hat{r}_e^k$) and previous($\hat{r}_e^{k-1}$) estimates. For $t \in[t_k,t_{k+1})$, the current reachable set is denoted by {\color{blue}$B_e\left(\hat{r}_e^k, \nu(t - t_k) + \gamma\right)$} and the previous reachable set is denoted by {\color{red}$B_e\left(\hat{r}_e^{k-1}, \nu(t - t_k + \hphiklg) + \gamma\right)$}. The actual reachable set is the intersection of the two.}
\label{fig:pe_analysis_multiple}
 \end{figure}

Consider the problem in which the pursuer receives uncertain information about the evader, while keeping track of its previous estimates. For the purpose of illustration, we analyze the case where the pursuer retains only the previous estimate of the evader's position. Extending the framework to the case for more than one previous estimates will be similar and straightforward. Additionally, we assume that the pursuer samples the evader with an associated error $\gamma$. 
Suppose that the pursuer sampled the evader at time $t_{k-1}$, computed the self-triggered update duration $\hphiklg$ and observed the evader again at the instance $t_k$. For $t \in [t_k, t_{k+1})$, the \emph{current} reachable set of the evader is given by $B_e\left(\hat{r}_e^k, \nu(t - t_k) + \gamma\right)$, and the \emph{previous} reachable set of the evader is given by $B_e\left(\hat{r}_e^{k-1}, \nu(t - t_k + \hphiklg) + \gamma\right)$. Based on this information, the \emph{actual} reachable set of the evader is given by the intersection of the two. One particular scenario is illustrated in Fig.~\ref{fig:pe_analysis_multiple}, where the actual reachable set is not the same as the current reachable set.

Using $\tau = t - t_k$ for $t \in [t_k, t_{k+1})$, let $\hat{g}_{\gamma}(\tau) = \underset{x_e, y_e, \theta_e}{\sup} \dot{R}_{\tau}(x_e,y_e,\theta_e)$, subject to the \emph{actual} reachable set of the evader i.e.
\[r_e \in B_e\left(\hat{r}_e^k, \nu\tau + \gamma\right) \cap B_e\left(\hat{r}_e^{k-1}, \nu(\tau + \hphiklg) + \gamma \right).\] The self-triggered update duration is then defined as
 \begin{align}\label{eq:self_trig_def_mult}
 \hphikg &= \inf \left\{\tau \in \realpositive \,|\, \hat{g}_{\gamma}(\tau) = 0 \right\}.
 \end{align}
Equivalently, for fixed $\tau$ and $\gamma$, $\hat{g}_{\gamma}(\tau)$ is the optimal value of the following optimization problem.
\begin{equation} \label{eq:opt_dot_R_2}
\begin{aligned}
& \underset{x_e, y_e, \theta_e}{\sup} &\dot{R}_{\tau}(x_e, y_e, \theta_e),\\
& \text{subject to} &r_e \in B_e\left(\hat{r}_e^{k-1}, \nu (\tau+ \hphiklg) + \gamma \right),\\
& & r_e \in B_e(\hat{r}_e^k, \nu \tau + \gamma).
\end{aligned}
\end{equation}

As explained earlier, incorporating evader's previous estimate can potentially reduce the \emph{actual} reachable set of the evader in the case of noisy measurements. By keeping track of the previous estimate(s), we are equivalently adding more constraints to the feasible (reachable) set in our optimization problem. We want to formalize the benefit of retaining evader's history in terms of improvement (increase) in the update duration, as compared to the memoryless update duration given by~\eqref{eq:self_trig_gamma}. Recall that the self-triggered update duration in~\eqref{eq:self_trig_gamma} was derived using only the current reachable set of the evader, $B_e(\hat{r}_e^k, \nu \tau + \gamma)$, based on its latest estimate. Let $g_{\gamma}(\tau)$ denote the optimal value of the problem~\eqref{eq:opt_dot_R_2_relax}, which is used to obtain the memoryless update duration in~\eqref{eq:self_trig_gamma}.
\begin{equation} \label{eq:opt_dot_R_2_relax}
\begin{aligned}
& \underset{x_e, y_e, \theta_e}{\sup} &\dot{R}_{\tau}(x_e, y_e, \theta_e),\\
& & r_e \in B_e(\hat{r}_e^k, \nu \tau + \gamma).
\end{aligned}
\end{equation}
Observe that~\eqref{eq:opt_dot_R_2_relax} is a relaxation of~\eqref{eq:opt_dot_R_2}, as it is obtained by removing the constraint corresponding to the previous reachable set of the evader. As a result, $\hat{g}_{\gamma}(\tau) \le g_{\gamma}(\tau)$ for any $\tau \ge 0$. We notice that both $\hat{g}_{\gamma}$ and $g_{\gamma}$ are monotonically increasing in the parameter $\tau$, because increasing $\tau$ increases the feasible set in the optimization problem thus yielding a potentially greater maximum value. Using the monotonicity of $\hat{g}$ and $g$ in $\tau$, along with the fact that $\hat{g}_{\gamma}(\tau) \le g_{\gamma}(\tau)$ (for any $\tau \ge 0)$, allows us to infer that the first instance, at which $\hat{g}$ approaches $0$, will be greater than or equal to the first instance at which $g$ approaches $0$. This means that
\[\inf \left\{\tau \in \realpositive\,|\, \hat{g}_{\gamma}(\tau) = 0 \right\} \ge \inf \left\{\tau \in \realpositive\,|\, g_{\gamma}(\tau) = 0 \right\}\]
and as a consequence, we have
\begin{align} \label{eq:prev_inequality}
\hphikg &\ge \phikg.
\end{align}

\paragraph*{Leveraging Memory against Uncertainty: }The relationship in~\eqref{eq:prev_inequality} shows that by using previous estimate(s) of the evader's position, we can potentially increase the update durations for the pursuer.
\begin{figure}[!h]
\centering
\begin{subfigure}[Case I]
{
  \includegraphics[scale = 0.3]{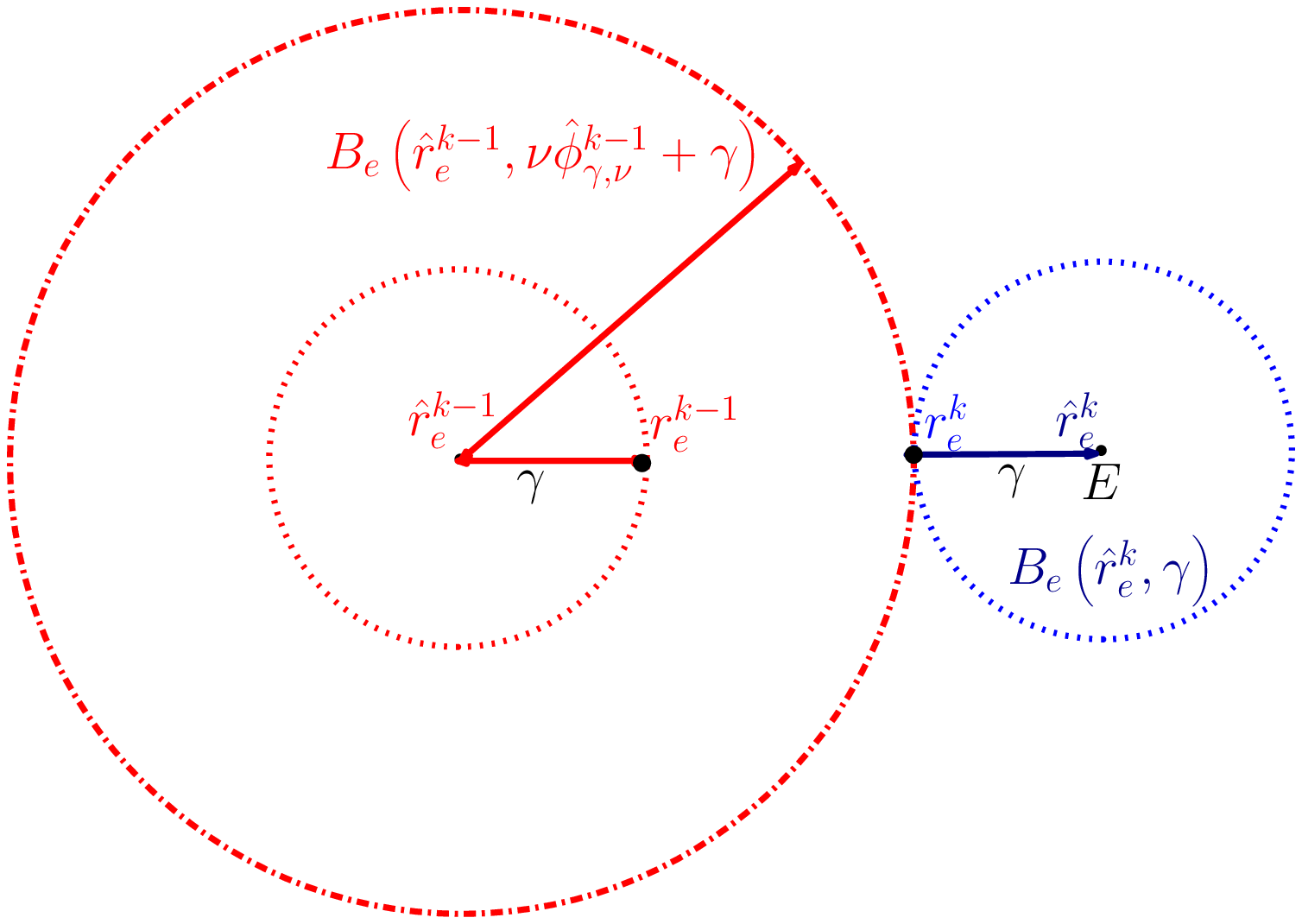}
  \label{fig:error_cases_1}}
\end{subfigure}%
\begin{subfigure}[Case II]
  {
  \includegraphics[scale = 0.3]{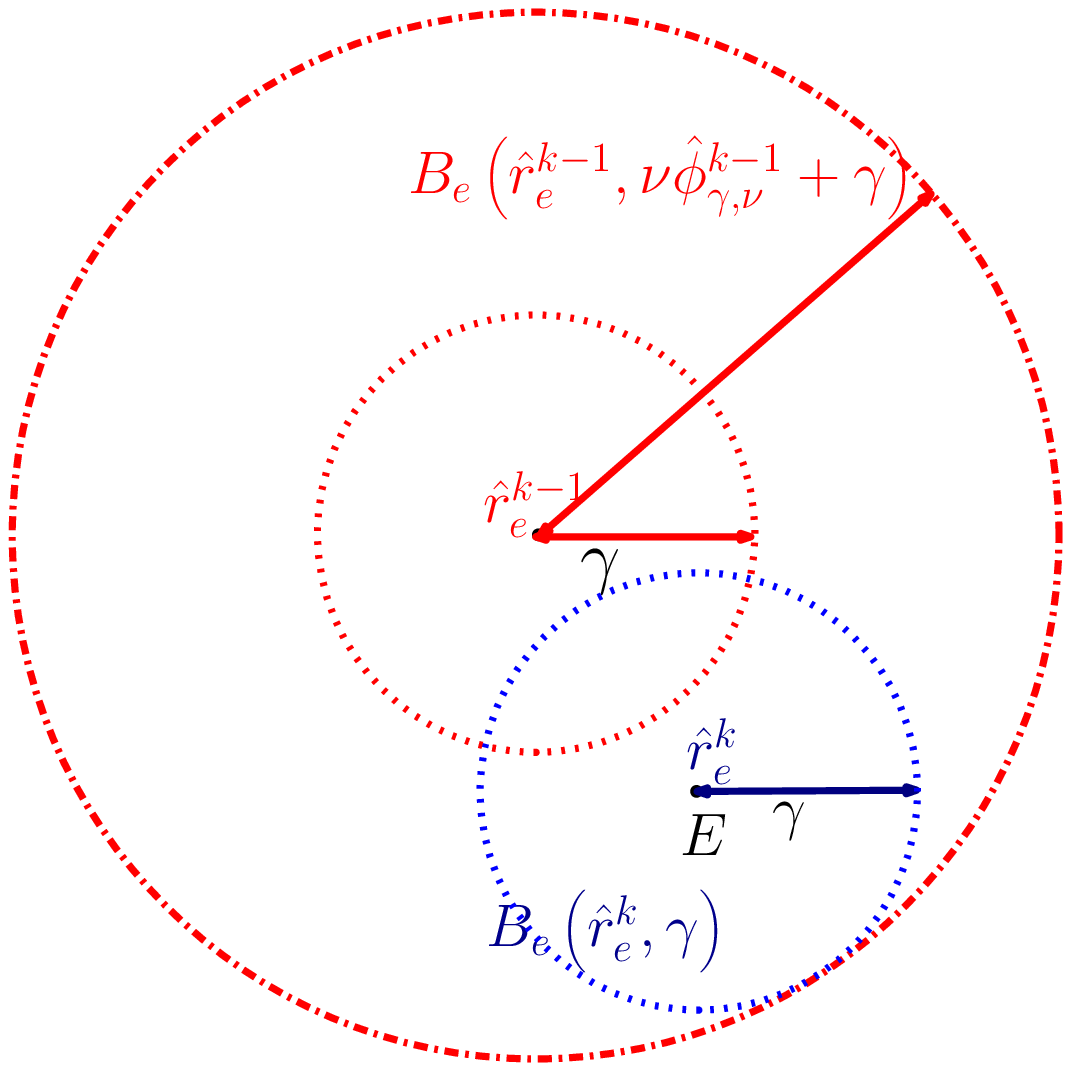}
  \label{fig:error_cases_2}}
\end{subfigure}
\caption{Figure shows the two possible cases at the instance of fresh observation of the evader. Case I shows the extreme case where incorporating the evader's history precisely determines its true position. Case II shows the scenario in which the current reachable set is a subset of the previous reachable set, thus the previous estimate provides no additional information.}
\label{fig:error_cases}
\end{figure}
Case I in Fig.~\ref{fig:error_cases} demonstrates one particular instance of leveraging evader's history which yields the greatest improvement in the current update duration. In the extreme case, at the instance of observation, the current and the previous reachable sets of the evader intersect such that the \emph{actual} reachable set is reduced to a point. Equivalently, this means that we know precisely where the evader is. This will result in a longer update duration as compared to memoryless case~\eqref{eq:self_trig_gamma}, where we would have incorporated uncertainty in our observation to yield a more conservative update duration. Let $\Delta \phi$ denote the improvement in self-triggered update duration, i.e. $\Delta \phi: = |\hphikg - \phikg|$. Comparing Eq.~\eqref{eq:t_min_v} and Eq.~\eqref{eq:self_trig_gamma}, we see that the greatest improvement (denoted by $\Delta \phi^{*}$) is given by
\begin{align}\label{eq:prev_improv}
\Delta \phi^{*} = \frac{2\gamma\left(\nu - \sqrt{1- \nu^2}\right)}{2\nu^2-1},
\end{align}
for $\nu \in [0,1)$.

While sometimes adding information about the previous estimate can be advantageous, it is important to realize that incorporating the previous reachable set of the evader does not always results in an improvement (increase) in the update duration. We can increase the self-triggered update duration only when the evader's past provides more information about its current true position. In the scenario where the current reachable set of the evader is a subset of the previous reachable set, we get no additional information about the evader's true position and hence no improvement in our update duration as compared to the memoryless case in~\eqref{eq:self_trig_gamma}. This is illustrated in case II in Fig.~\ref{fig:error_cases}, where we can drop (forget) the previous estimate of the evader's position as it is a subset of the previous reachable set and will yield no improvement in increasing the current update duration.

\begin{remark}[Numerical Example]
  \rm To illustrate the improvement in update duration numerically, let $\nu = 0.5$ and $\gamma = 0.1$. For these values, $\Delta \phi^{*} = 0.146$ from~\eqref{eq:prev_improv}. Based on some initial measurement, the pursuer finds the first update duration $\phi_{\gamma,\nu}^0$ by using~\eqref{eq:self_trig_gamma} and samples the evader at $t_1$ to find that the previous and the current reachable set intersect at a point. Suppose the measured separation at $t_1$ is $\hat{D}_1 = 10$ units. Using a memoryless pursuit policy in~\eqref{eq:self_trig_gamma} results in $\phi_{\gamma,\nu}^1 = 6.529$, whereas obtaining self-triggered update duration from solving the problem~\eqref{eq:opt_dot_R_2} results in $\hat{\phi}_{\gamma,\nu}^1 = 6.674$. Observe that $\hat{\phi}_{\gamma,\nu}^1 - \phi_{\gamma,\nu}^1 \approx \Delta \phi^{*}$, which shows that, in certain cases, knowing one previous estimate of the evader can almost nullify the uncertainty in sensing evader's position and consequently allow for greater update duration.
  \oprocend
\end{remark}

\paragraph*{Towards Multiple Estimates:} Extending the above framework to the case of multiple previous estimates is relatively straightforward. Suppose, for the $k$-th observation of the evader, we have the information about the $m$ previous updates (where $m \in \integernonnegative$ such that $m \le k - 1$). Then the current sleep duration can be computed by $\hphikg  = \inf \left\{\tau \in \realpositive \,|\, \hat{g}_{\gamma}(\tau) = 0 \right\}$, where $\hat{g}_{\gamma}(\tau)$ is an optimal value of the optimization problem~\eqref{eq:opt_dot_R_m} for $\tau \ge 0$. The value $m$ can be treated as the length of the sliding window for retaining fixed number of previous estimates, to compute the current update duration for the pursuer.
\begin{equation} \label{eq:opt_dot_R_m}
\begin{aligned}
& \underset{x_e, y_e, \theta_e}{\sup} &\dot{R}_{\tau}(x_e, y_e, \theta_e),\\
& \text{subject to} & r_e \in B_e(\hat{r}_e^k, \nu \tau + \gamma),\\
& & r_e \in \bigcap_{j=1}^m B_e^j(\tau),
\end{aligned}
\end{equation}
where $B_e^j(\tau)$ is the current reachable set of the $j$-th previous estimate of the evader's position and is given by
   \[B_e^j(\tau):= B_e\left(\hat{r}_e^{k-j}, \nu \left(\tau + \sum_{l = 1}^{j}\hphikpg\right) + \gamma \right),\]
for $j \in \{1, \ldots, m\}$ and $m \le k - 1$.

\begin{remark}[Forgetting Previous Estimates]
{\rm
Note that, while we might have the capability to store $m$ previous estimates of the evader, it is not necessary to use all of them in computing the current sleep duration for the pursuer. As illustrated earlier, retaining the history improves our update duration when it reduces the current reachable set of the evader. This allows us to forget all those estimates whose reachable sets either completely contain the current reachable set or a reachable set of another previous estimate. To formalize this notion, at the instance of $k$-th observation of the evader, we can construct a set

\[ \mathcal{I} := \left\{ i\, |\, B_e\left(\hat{r}_e^k,\gamma\right) \subseteq B_e^i(0)\right\} \cup \left\{ i\, |\, B_e^l(0)\subseteq B_e^i(0), \, l \ne i\right\},\]
where $B_e^i(0)$ denotes $B_e\left(\hat{r}_e^{k-i}, \nu \sum_{l = 1}^{i}\hphikpg + \gamma \right)$ for $i \in \{1, \ldots, m\}$ and $m \le k - 1$. $\mathcal{I}$ denotes the collection of indices among the $m$ previous samples of all those estimates which we can forget to reduce the computation complexity of the problem~\eqref{eq:opt_dot_R_m}. Thus, our improved update duration can be computed from solving the problem~\eqref{eq:opt_dot_R_m_I}. The problem~\eqref{eq:opt_dot_R_m_I} is guaranteed to have the same optimal value as that of~\eqref{eq:opt_dot_R_m} because all the estimates belonging to the set $\mathcal{I}$ have no effective contribution to the actual reachable set of the evader and thus removing them will have no change in the optimal value.
\begin{equation} \label{eq:opt_dot_R_m_I}
\begin{aligned}
& \underset{x_e,y_e,\theta_e}{\sup} &\dot{R}_{\tau}(x_e, y_e, \theta_e),\\
& \text{subject to} & r_e \in B_e(\hat{r}_e^k, \nu \tau + \gamma),\\
& & r_e \in \bigcap_{j=1, j \notin  \mathcal{I}}^{m} B_e^j(\tau).
                \end{aligned}
\end{equation}
                }\oprocend

              \end{remark}

\section{Simulations}\label{se:simulations}
In this section, we provide numerical results for the case when the pursuer retains past samples of the evader as outlined in Section~\ref{se:prev_estimates}. We study the potential benefit of retaining only the previous estimate of the evader's observation as the pursuer tries to capture the evader. Thus, in our simulations, $m = 1$ in~\eqref{eq:opt_dot_R_m_I} and the self-triggered update duration will be obtained from solving the optimization problem in~\eqref{eq:opt_dot_R_2}.

We model our agents as single integrators, where we normalize the speeds of the agents, such that $v_p = 1$ and the maximum speed of the evader is $\nu = 0.5$. The evader is restricted to move in any of $4$ directions: right, left, up and down and chooses the best direction to actively move away, with its maximum possible speed, from the pursuer at all times. We initialize the agents with an actual separation $D_0 = 15$ units. For every observation, the pursuer samples the evader's current position with an associated error of $\gamma = 0.1$ such that, initially, the true position of the evader $r_e^k \in B_e\left(\hat{r}_e^k,0.1\right)$. Note that we can not arbitrarily set the capture radius, as outlined in Section~\ref{se:error}. For an evader speed $\nu$ and error $\gamma$, the capture radius must satisfy the relationship $\epsilon > \frac{\gamma}{p(\nu)}$, where $p(\nu) = \frac{(1 - \nu)\sqrt{1 - \nu^2} - 2(1 - \nu)^2}{5\nu - 3}$. For $\nu = 0.5$ and $\gamma = 0.1$, setting $\epsilon = 0.75$ satisfies the aforementioned relationship. Table $1$ shows the variation between the \emph{memoryless} update duration ${\phi}_{\gamma,\nu}^k$ and the improved update duration $\hat{\phi}_{\gamma,\nu}^k$, which takes into account the previous estimate of the evader's observation. Since, $\hat{\phi}_{\gamma,\nu}^k$ is different from ${\phi}_{\gamma,\nu}^k$, they result in different measures of separation at the instance of observation. For a fair comparison, we need to compare the normalized update durations, i.e $\frac{\hat{\phi}_{\gamma,\nu}^k}{\bar{D}_k}$ and $\frac{\phi_{\gamma,\nu}^k}{D_k}$ where $\bar{D}_k$ and $D_k$ denote the true separations at the instance of evader observation for memory-aware and memoryless pursuit strategies, respectively. The results in Table~\ref{table:prev_hist_1} indicate that in the presence of uncertainty in sampling the evader, incorporating only the previous estimate allows for greater sleep durations that guarantees capture with finite updates. We observe that, in the beginning of the pursuit, adding history results in relatively better gains as compared to towards the end, when the agents are nearby. The values indicate that as the separation decreases, so does the potential benefit of adding computational overhead by retaining the previous observation.

\begin{table}

\begin{center}
 \begin{tabular}{||c||c|c|c||c|c|c||}
\hline
\textbf{$k$}&\textbf{$D_k$}&\textbf{$\phi_{\gamma,\nu}^k$}&\color{blue}\textbf{$\frac{\phi_{\gamma,\nu}^k}{D_k}$}
&\textbf{$\bar{D}_k$}&\textbf{$\hat{\phi}_{\gamma,\nu}^k$}&\color{red}\textbf{$\frac{\hat{\phi}_{\gamma,\nu}^k}{\bar{D}_k}$}\\ \hline
\textbf{0}&15.00&    9.500&\color{blue}    0.633&   15.00&    9.500&\color{red}    0.633\\ \hline
\textbf{1}&10.25&    6.488&\color{blue}    0.633&   10.24&    6.674&\color{red}    0.652\\ \hline
\textbf{2}&7.006&    4.432&\color{blue}    0.633&    6.949&    4.511&\color{red}    0.649\\ \hline
\textbf{3}&4.790&    3.027&\color{blue}    0.632&    4.659&    2.996&\color{red}    0.643\\ \hline
\textbf{4}& 3.277&    2.067&\color{blue}   0.631&    3.196&    2.037&\color{red}    0.637\\ \hline
\textbf{5}& 2.243&   1.412&\color{blue}   0.630&    2.194&    1.393&\color{red}    0.634\\ \hline
\textbf{6}&1.537&    0.965&\color{blue}    0.628&    1.500&    0.947&\color{red}    0.632\\ \hline
\textbf{7}&1.055&    0.659&\color{blue}    0.625&    1.033&    0.648&\color{red}    0.628\\ \hline
\textbf{8}&0.754&   0.463&\color{blue}    0.621&    0.752&    0.464&\color{red}    0.623\\ \hline

\end{tabular}
\caption{Comparison between memoryless and memory-aware update times.}
\label{table:prev_hist_1}
\end{center}

\end{table}
\section{Conclusions}\label{se:conclusions}
 The robust self-triggered framework in this paper extends our previous results to address the practical issues related with uncertainty in information about the evader. We elaborate the case when the sampling is not perfect and design self-triggered update duration along with tolerable error bounds in estimating the evader's state. We show that our analysis preserve the self-triggered controller updates for the pursuer, such that it incurs no Zeno behavior in catching the evader without losing any of the previous performance guarantees. Our methodology offers a fresh perspective on dealing with uncertain information in pursuit-evasion problems, besides being in contrast to a majority of previous works that assume continuous, or at least periodic, information about the evader is available at all times. Additionally, we study the merits of retaining evader's history and show that we can allow for potentially longer update durations by incorporating past observations of the evader in the pursuer's autonomous decision making. In the future, we are interested in extending our methods to scenarios involving multiple agents and deriving conditions for team-triggered cooperative strategies.



{\scriptsize
\bibliographystyle{plainnat}
\bibliography{root}
}

\appendix
\section{Derivation of Self-triggered Update Duration}\label{se:appendix}

\begin{figure}[!h]
\centering
\includegraphics[scale =.55]{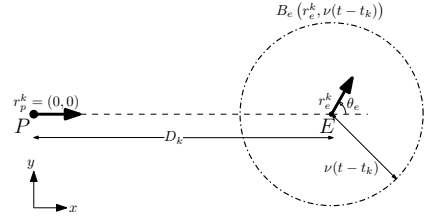}
\caption{Figure shows the pursuer and the evader, at $r_p^k = (0,0)$ and $r_e^k = (D_k, 0)$ respectively, separated by $D_k$ at time $t_k$. $B_e\left(r_e^k, \nu (t-t_k)\right)$ is the ball centered at $r_e^k$ with radius $\nu (t-t_k)$ and indicates the reachable set of the evader for $t \in [t_k, t_{k+1})$.}
\label{fig:pe_analysis}
\end{figure}

The agents are modelled by the dynamics in~\eqref{eq:pe_dyn}. At time $t_k$, the pursuer observes the evader at a distance $D_k = \|r_e^k - r_p^k\|$. Without loss of generality, we make the relative vector between pursuer and the evader parallel to the $x-$axis such that $y_p(t_k) = 0$ and $y_e(t_k) = 0$.  Additionally, as a matter of convenience, we assume that $0 = x_p(t_k) < x_e(t_k) = D_k$. This is elaborated in Fig.~\ref{fig:pe_analysis}.

Thus, our pursuit trajectory is parallel to the $x-$axis as $\theta_p(t_k) = 0$. The pursuer does not observe the evader till the next update instance, thus $\theta_p(t) = 0$ for $t_{k+1} - t_k$. The modified dynamics are given by
\begin{equation}\label{eq:pe_mod_dyn}
\begin{aligned}
\dot{x}_p &= 1, &\dot{x}_e &= \nu \cos \theta_e,\\
\dot{y}_p &= 0, &\dot{y}_e &= \nu \sin \theta_e.
\end{aligned}
\end{equation}
From~\eqref{eq:pe_mod_dyn}, $r_p(t) = (t-t_k,0)$ for $t \in [t_k,t_{k+1})$. Thus, $\dot{R} = (x_e-x_p)(\dot{x}_e - \dot{x}_p) + (y_e-y_p)(\dot{y}_e - \dot{y}_p)$, where $R = \frac{r^2}{2}$ and $r$ is the separation. $\dot{R} = \nu(x_e - t + t_k) \cos \theta_e + \nu y_e \sin \theta_e + t - t_k - x_e.$ Using $\tau \triangleq t - t_k$,
\begin{align}\label{eq:dot_R}
\dot{R} &= \nu(x_e - \tau) \cos \theta_e + \nu y_e \sin \theta_e + \tau - x_e.
\end{align}

The agent updates when $\dot{R}$ can possibly become nonnegative. For $\tau \ge 0$, $r_e \in B_e(r_e^k, \nu \tau)$. $\dot{R}_{\tau}(x_e, y_e, \theta_e)$ explicitly denotes $\dot{R}$ in~\eqref{eq:dot_R} in terms of evader parameters, for fixed $\tau$. The problem is formulated as
\begin{align}\label{eq:opt_dot_R}
& \underset{x_e, y_e, \theta_e}{\sup}
& &  \dot{R}_{\tau}(x_e, y_e, \theta_e),\\
& \text{subject to}
& & (x_e,y_e) \in B_e(r_e^k, \nu \tau) . \notag
\end{align}
For $\tau \ge 0$, $g(\tau)$ denotes the optimal value of problem~\eqref{eq:opt_dot_R} and the self-triggered update duration is defined as
 \[\phik = \inf \left\{\tau \in \realpositive \,|\, g(\tau) = 0 \right\}.\]

In $\real^2$, $B_e(r_e^k, \nu \tau) \triangleq (x_e - D_k)^2 + y_e^2 \le (\nu \tau)^2$. The constraint of the problem~\eqref{eq:opt_dot_R} is independent of $\theta_e$. $\frac{\partial \dot{R}_{\tau}}{\partial \theta_e} = -\nu (x_e - \tau)\sin \theta_e +\nu y_e \cos \theta_e$. Setting $\frac{\partial \dot{R}_{\tau}}{\partial \theta_e} = 0$ yields $\theta_e^* = \arctan\left( \frac{y_e}{x_e - \tau}\right)$.
Note that for $\tau \in [0,\phik )$, $x_e - \tau \ge 0$. To see this, suppose $x_e - \tau < 0$. Setting $\theta_e = 0$ in~\eqref{eq:dot_R}, we get $\dot{R}_{\tau} = -(1 - \nu)(x_e - \tau) > 0$ for $\nu \in [0,1)$, which is a contradiction as $\dot{R} \le 0$ for any $\tau \in [0,\phik )$. Due to symmetry of the problem, we can assume $y_e \ge 0$. Since $x_e - \tau \ge 0$, we have $\theta_e^* \in [0, \frac{\pi}{2}]$. Substituting $\theta_e^*$ in~\eqref{eq:dot_R}, we get
\[
\dot{R}_{\tau}(x_e,y_e) = \nu {\sqrt{y_e^2 + (x_e - \tau)^2}} + \tau - x_e,\]
which simplifies the problem in~\eqref{eq:opt_dot_R} to
\begin{align}\label{eq:opt_dot_R_xy}
& \underset{x_e, y_e}{\sup}
& &  \dot{R}_{\tau}(x_e, y_e) = \nu {\sqrt{y_e^2 + (x_e - \tau)^2}} + \tau - x_e\\
& \text{subject to}
& & (x_e,y_e) \in B_e(r_e^k, \nu \tau).\notag
\end{align}
From the constraint $(x_e,y_e) \in B_e(r_e^k, \nu \tau)$ we get $y_e^2 \le (\nu \tau)^2 -(x_e - D_k)^2$,
which means that $y_e^*$ must lie at the boundary of $B_e(r_e^k, \nu \tau)$. Thus, $y_e^* = \sqrt{(\nu \tau)^2 -(x_e - D_k)^2}.$
Substituting $y_e^*$ in $\dot{R}_{\tau}(x_e,y_e)$, we get \[\dot{R}_{\tau}(x_e) = \nu {\sqrt{(x_e - \tau)^2 + (\nu \tau)^2- (x_e - D_k)^2}}+ \tau - x_e.\]
This reduces the problem in~\eqref{eq:opt_dot_R_xy} to
\begin{align}\label{eq:opt_dot_R_x}
 &\underset{x_e}{\sup} & & \dot{R}_{\tau}(x_e) \\
 &\text{subject to} & & x_e  \in [D_k - \nu \tau, D_k + \nu \tau]. \notag
\end{align}

Note that we can relax the problem in~\eqref{eq:opt_dot_R_x} by omitting the constraint $x_e  \in [D_k - \nu \tau, D_k + \nu \tau]$. The relaxation of~\eqref{eq:opt_dot_R_x} results in an unconstrained optimization problem. Ignoring the constraints of problem~\eqref{eq:opt_dot_R_x}, let $\tilde{g}(\tau) \triangleq \underset{x_e}{\sup}\, \dot{R}_{\tau}(x_e)$ and $\tilde{\phi}_{\nu}^k = \inf \{\tau \in \realpositive\,|\, \tilde{g}(\tau) = 0\}$. As a result of this relaxation, we have $\tilde{\phi}_{\nu}^k  \le \phik$. Let $\tilde{x}_e^*:= \argmax\, \dot{R}_{\tau}(x_e)$ for the unconstrained problem. To perform unconstrained maximization of $\dot{R}_{\tau}(x_e)$, the derivative is given by $\frac{\partial \dot{R}_{\tau}}{\partial x_e} = \frac{\nu (D_k - \tau)}{\sqrt{(x_e - \tau)^2 + (\nu \tau)^2- (x_e - D_k)^2}} - 1$.
Setting $\frac{\partial \dot{R}_{\tau}}{\partial x_e} =0$ yields
$\tilde{x}_e^* = \frac{D_k^2\nu^2 + D_k^2 - 2\tau D_k\nu^2 - \tau^2}{2(D_k- \tau)}$. As $\tilde{x}_e^*:= \argmax\, \dot{R}_{\tau}(x_e)$, $\tilde{g}(\tau) = \dot{R}_{\tau}(\tilde{x}_e^*)$ and is given by $\tilde{g}(\tau) = \tau - \frac{D_k^2\nu^2 + D_k^2 - 2\tau D_k\nu^2 - \tau^2}{2(D_k - \tau)} + \nu^2(D_k - \tau)$.
Solving for $\inf\{ \tau \in \realpositive\,|,\ \tilde{g}(\tau) = 0\}$, yields
\begin{align}\label{eq:t_min}
\tilde{\phi}_{\nu}^k &= \frac{D_k\nu \sqrt{1 - \nu^2} - D_k(1 - \nu^2)}{2\nu^2 - 1}.
\end{align}
Recall that $\tilde{\phi}_{\nu}^k \le \phik$, where $\tilde{\phi}_{\nu}^k$ was obtained from relaxing the constraint in problem~\eqref{eq:opt_dot_R_x}. Our claim is that at the instance of update ($\tau = \tilde{\phi}_{\nu}^k$), the maximizer $\tilde{x}_e^*$ of the relaxed problem is a feasible solution of the problem~\eqref{eq:opt_dot_R_x}, i.e. $\tilde{x}_e^*(\tau) \in [D_k - \nu \tau, D_k + \nu \tau]$ for $\tau = \tilde{\phi}_{\nu}^k$. To see this, at $\tau = \tilde{\phi}_{\nu}^k$, the maximizer $\tilde{x}_e^*$, $D_k + \nu\tau$ and $D_k - \nu\tau$ are given by
\begin{align*}
D_k - \nu\tilde{\phi}_{\nu}^k &= D_k\varphi_{-}(\nu),\\
\tilde{x}_e^*(\tilde{\phi}_{\nu}^k) &= D_k\varphi_2(\nu),\\
D_k + \nu\tilde{\phi}_{\nu}^k&= D_k\varphi_{+}(\nu),
\end{align*}
where \begin{align*}
\varphi_{\pm}(\nu) &= 1 \pm \frac{\nu^2 \sqrt{1 - \nu^2} - \nu(1 - \nu^2)}{2\nu^2 - 1},\\
\varphi_2(\nu) &= \frac{3\nu^3 + \sqrt{1 - \nu^2} -2\nu - 2\nu^4\sqrt{1 - \nu^2} }{(\nu -\sqrt{ 1 - \nu^2})(2\nu^2 - 1)}.
\end{align*}
For $ \nu \in [0,1)$, $\varphi_{-}(\nu) \le \varphi_2(\nu) \le \varphi_{+}(\nu)$. This shows that $\tilde{x}_e^*(\tilde{\phi}_{\nu}^k)$ satisfies the constraints in the problem~\eqref{eq:opt_dot_R_x} at $\tau = \tilde{\phi}_{\nu}^k$. This means, at the instance of update, the maximizer $\tilde{x}_e^*$ of the relaxed problem is a feasible solution of the original problem~\eqref{eq:opt_dot_R_x} and hence it is optimal solution $x_e^*$ for~\eqref{eq:opt_dot_R_x}. Thus $\tilde{g}(\tau) = g(\tau)$ and as a result $\tilde{\phi}_{\nu}^k = \phik$. Note that $\phik$ is continuous in the parameter $\nu$ as $\lim_{\nu \to \frac{1}{\sqrt{2}}} \frac{D_k\nu \sqrt{1 - \nu^2} - D_k(1 - \nu^2)}{2\nu^2 - 1} = \frac{D_k}{2}$.

\section{Maximum Separation}
\begin{figure}[!h]
\centering
\includegraphics[scale =.65]{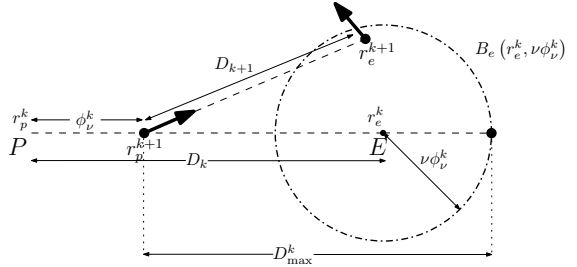}
\caption{$D_{k+1}$ indicates the new separation between the agents at $t_{k+1}$. $B_e(r_e^k, \nu \phik)$ outlines the boundary of the reachable set of the evader after $\phik$. $D_{\max}^k$ denotes the maximum possible separation between the pursuer and the evader at $t = t_k + \phik$.}
\label{fig:pe_analysis_cont}
\end{figure}

If the pursuer updates its trajectory using the self-triggered update policy described in~\eqref{eq:t_min}, then the maximum distance between the agents, between successive updates, is given by
\begin{align*}
D_{\max}^k = D_k - (1 - \nu)\phik.
\end{align*}
To see this, after the duration $\phik$, the pursuer moves a distance of $\phik$ units and the evader evader can be anywhere inside a ball of radius $\nu \phik$ centered at $r_e^k$. This is shown in Fig.~\ref{fig:pe_analysis_cont}. The maximum separation between the pursuer and the evader is denoted by $D_{\max}^k$  and is given by $D_k - (1 - \nu)\phik$.

\end{document}